\newcommand{\bashinline}[1]{\texttt{\detokenize{#1}}}
\documentclass[fleqn,usenatbib]{rasti}

\usepackage{newtxtext,newtxmath}

\usepackage[T1]{fontenc}

\DeclareRobustCommand{\VAN}[3]{#2}
\let\VANthebibliography\thebibliography
\def\thebibliography{\DeclareRobustCommand{\VAN}[3]{##3}\VANthebibliography}

\usepackage{graphicx}	
\usepackage{amsmath}	
\usepackage[table]{xcolor}	
\usepackage[most]{tcolorbox}
\usepackage{algorithm}
\usepackage{algpseudocode}
\usepackage{float}

\title[O Mirror, Where Art Thou?]{O Mirror, Where Art Thou? \\
  \large Position retrieval using nested sampling and global minimization}

\author[Sai Krishanth P.M. et al.]{
Sai Krishanth P.M.,$^{1}$\thanks{E-mail: saikrishanth@arizona.edu}
Sanchit Sabhlok,$^{1}$
Jaren N. Ashcraft,$^{2}$
Kyle Van Gorkom,$^{1}$
and Ewan S. Douglas$^{1}$\thanks{E-mail: douglase@arizona.edu}
\\
$^{1}$Steward Observatory\\
$^{2}$UC Santa Barbara
}

\date{Accepted XXX. Received YYY; in original form ZZZ}

\pubyear{\the\year{}}

\begin{document}
\label{firstpage}
\pagerange{\pageref{firstpage}--\pageref{lastpage}}
\maketitle

\begin{abstract}
Wavefront sensing reveals that a telescope is misaligned, but not which mirror moved or by how much; inferring the mirror positions themselves turns alignment into a direct correction. We describe a simulation framework for inferring telescope mirror misalignment from multi-field Point Spread Function (PSF) measurements projected onto Circular-Bessel and Zernike basis vectors. Two case studies are presented: a Hubble Space Telescope model with five secondary-mirror degrees of freedom and a three-mirror anastigmat with ten primary and secondary mirror degrees of freedom. In both cases, field-dependent multivariate Gaussian likelihoods are explored with nested sampling and refined with differential evolution. Recovered states are validated with residual PSF images and optical path difference (OPD) measurements. In the Hubble benchmark, differential evolution recovers injected misalignments to $\lesssim 0.01\,\mu\mathrm{m}$ and reduces on-axis OPD error from $\sim 2.3$ waves to $\sim 3\times 10^{-8}$ waves. In the TMA study, the posterior shows broad marginals and correlated parameter pairs that suggest partial degeneracies at higher dimension; differential evolution recovers the truth alignment state less precisely than in the Hubble benchmark ($\lesssim 17\,\mu\mathrm{m}$) and reduces on-axis OPD error from $\sim 8.24$ waves to $\sim 2\times 10^{-3}$ waves. This generalizable algorithm enables direct sensing of the state of an optical system and automated alignment.
\end{abstract}




\section{Introduction}\label{sec:intro}

Optical observatories depend on precise alignment of telescope optics to maintain image quality. Wavefront aberrations from mirror translations and tilts propagate into field-dependent point spread functions (PSFs). These field-dependent PSFs therefore provide observables sensitive to rigid-body misalignment even when direct wavefront sensing is unavailable or incomplete.

In this work, we implement a simulation-based solution to the inverse problem of estimating the wavefront from observables: given multi-field PSF observations projected onto basis vectors, we estimate the underlying telescope mirror alignment parameters $\boldsymbol{\theta}$. Synthetic PSFs generated from an open-source ray-tracing model are compared to stored ``observations'' through a multivariate Gaussian likelihood in coefficient space. We present benchmarks against example two and three mirror telescope models as summarized in Table~\ref{tab:summary}.

\begin{table*}
\centering
\caption{Summary of the two telescope models used in this study.}
\label{tab:summary}
\begin{tabular}{llllll}
\hline
Telescope & Degrees of freedom & Noise & Basis (modes) & Fields & Wavelength \\
\hline
Hubble & 5  & No  & Circular-Bessel (15) & 5  & 0.55\,$\mu$m \\
TMA    & 10 & Yes & Zernike (11) & 10 & 0.65\,$\mu$m \\
\hline
\end{tabular}
\end{table*}

Historical HST servicing relied on phase retrieval with focus diversity~\citep{Fienup1982,Lofgren2014,Krist1997} rather than direct PSF-to-alignment inversion leveraging a full ray-trace model. The commissioning of the James Webb Space Telescope (JWST),  a deployable three-mirror anastigmat~\citep{Knight2012}, is the most relevant recent precedent. Because the segmented primary can compensate the on-axis wavefront error caused by a misplaced secondary mirror, secondary-mirror misalignment appears mainly in field-dependent terms (astigmatism and focal-plane tilt) that can only be sensed at multiple field points~\citep{ActonKnight2012}. The final stage of JWST telescope commissioning, Multi-Instrument Multi-Field (MIMF) alignment, measured the wavefront with phase retrieval at field points across the science instruments to verify and refine the secondary-mirror position~\citep{Acton2022,McElwain2023}. That approach reconstructs the wavefront at each field point and maps it to mirror corrections through a linear optical model; here we instead infer rigid-body mirror positions directly from PSF intensity with a non-linear ray-trace forward model, obtaining a posterior over the alignment parameters.

Modern open-source ray tracers and nested samplers make it practical to evaluate a full forward model at every likelihood call over a large enough parameter space for the method to be practically useful. Missions such as the Habitable Worlds Observatory (HWO;~\citealt{HWO2023}) and the Lazuli Space Observatory~\citep{Roy2026} will require precise and robust alignment algorithms for complex three-mirror anastigmat designs. We investigate the feasibility of solving for telescope misalignments directly from PSF observables as a method for future observatories, like HWO, to rapidly and precisely stabilize their mirrors. 

\section{Methods}

Misalignments produce aberrations, but phase retrieval recovers only the wavefront; inferring the positions of the telescope optics requires a second, model-based step, such as the linear multi-field inversion used during JWST commissioning (Section~\ref{sec:intro}). In this work, we instead use a ray-trace model of the telescope to infer mirror misalignments directly from PSF observations.

\begin{figure*}
    \centering
    \includegraphics[width=\textwidth]{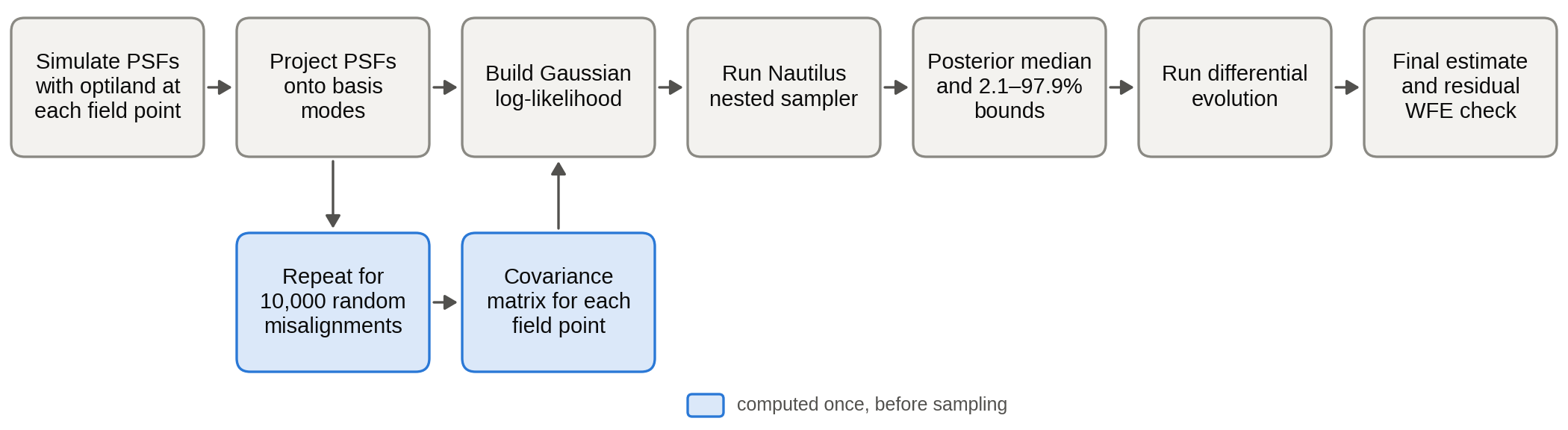}
    \caption{Inference pipeline: Optiland forward model, basis-coefficient feature extraction, multivariate Gaussian likelihood, Nautilus nested sampling, and differential-evolution refinement with PSF/OPD validation.}
    \label{fig:block}
\end{figure*}

To perform misalignment inference, we construct a pipeline (Figure~\ref{fig:block}) that does the following:
\begin{enumerate}
    \item Build a parametric telescope in Optiland~\citep{Harrison_Optiland_2026} and propagate monochromatic light to generate a synthetic PSF.
    \item Project PSF onto a basis.
    \item Model observations as multivariate Gaussians in coefficient space with field-specific covariances $\boldsymbol{\Sigma}_f$ from Monte Carlo calibration.
    \item Explore the posterior with Nautilus~\citep{Lange2023}, and then refine a point estimate with differential evolution.
    \item Validate by comparing nominal and residual-alignment systems in PSF and OPD space.
\end{enumerate}

\subsection{Nested sampling}
Nested sampling~\citep{Skilling2006} is a Bayesian Monte Carlo method that estimates the evidence $\mathcal{Z}=\int \mathcal{L}(\boldsymbol{\theta})\pi(\boldsymbol{\theta})\,\mathrm{d}\boldsymbol{\theta}$ (where $\mathcal{L}(\boldsymbol{\theta})$ is the likelihood and $\pi(\boldsymbol{\theta})$ is the prior) while producing weighted posterior samples. A set of $K$ live points is maintained in prior space and the point with the lowest likelihood is iteratively replaced by a new draw from the prior subject to a rising likelihood threshold. Discarded points contribute to $\ln\mathcal{Z}$ and  surviving points approximate $\pi(\boldsymbol{\theta}\mid\mathrm{data})$.

We use \texttt{Nautilus}, which implements importance nested sampling~\citep{Feroz_2019} with neural-network proposals for efficient exploration of the prior. The prior is encoded by a transform \bashinline{ptform} that maps uniform unit-cube coordinates $\mathbf{u}\in[0,1]^d$ to physical box bounds on each component of $\boldsymbol{\theta}$. The function passed to the sampler returns $\ln\mathcal{L}(\boldsymbol{\theta})$ only; the prior density is uniform in the box and zero elsewhere. A high-level description of nested sampling is presented in Algorithm~\ref{alg:cap}.

\begin{algorithm}
\caption{Nested sampling algorithm}\label{alg:cap}
\begin{algorithmic}
\State Draw $K$ live points from the prior
\While{stopping criterion not met}
\State Compute the minimum likelihood $L_{\min}$ from the current set of live points
\State Add the $k^{\mathrm{th}}$ live point associated with $L_{\min}$ to a list of dead points
\State Sample a new point from the prior with likelihood $> L_{\min}$
\State Replace the $k^{\mathrm{th}}$ live point with the newly sampled point
\State Evaluate stopping criterion
\EndWhile
\State Sequentially discard remaining live points to complete the evidence estimate
\end{algorithmic}
\end{algorithm}

The log-likelihood sums independent multivariate Gaussian likelihoods over the normalized field index $f\in\mathcal{F}$. For the TMA, $\mathcal{F} = \{(0,0.5965),\allowbreak(1,0),\allowbreak(0,1),\allowbreak(1,1),\allowbreak(0,-1),\allowbreak(-1,0),\allowbreak(1,-1),\allowbreak(-1,1),\allowbreak(-1,-1),\allowbreak(0.5,0.5)\}$, where the near-axis field $(0,0.5965)$ is used in place of $(0,0)$; the Hubble field set is given in Section~\ref{sec:hubble}. The log-likelihood is given by equation~(\ref{eq:loglike}):
\begin{equation}
\begin{split}
\ln \mathcal{L}(\boldsymbol{\theta})
= \sum_{f \in \mathcal{F}} \Bigl\{
&-\frac{1}{2}
\left[\mathbf{c}_f(\boldsymbol{\theta}) - \mathbf{c}_{f,\mathrm{obs}}\right]^{\!\top}
\boldsymbol{\Sigma}_f^{-1}
\left[\mathbf{c}_f(\boldsymbol{\theta}) - \mathbf{c}_{f,\mathrm{obs}}\right] \\
&-\frac{N_z}{2}\ln(2\pi)
-\frac{1}{2}\ln\det\boldsymbol{\Sigma}_f
\Bigr\}
\end{split}
\label{eq:loglike}
\end{equation}
where $\mathbf{c}_f(\boldsymbol{\theta})$ is the forward-model basis-coefficient vector at field $f$, $\mathbf{c}_{f,\mathrm{obs}}$ is the synthetic observation, and $N_z$ is the number of retained modes (15 for Hubble, 11 for the TMA), and $\boldsymbol{\Sigma}_f$ is the field-specific covariance matrix.

\subsection{Differential evolution}
Differential evolution (DE) is a population-based global optimizer that refines a \texttt{Nautilus}-derived starting point into a single best fit misalignment vector by minimizing coefficient-space residuals. It was introduced by \citet{StornPrice1997} and is implemented in  \bashinline{scipy.optimize} as \bashinline{differential_evolution}.
DE maintains a population of candidate parameter vectors $\{\theta_1, \dots, \theta_N\}$ inside box bounds. For each generation, it iteratively improves the candidate solution as follows:
\begin{enumerate}
    \item Pick a target individual $\theta_i$.
    \item Build a trial vector $\theta_{\mathrm{trial}}$ by combining traits from other population members.
    \item Evaluate the objective at $\theta_{\mathrm{trial}}$.
    \item Keep $\theta_{\mathrm{trial}}$ if it is better than $\theta_\mathrm{i}$; otherwise keep $\theta_\mathrm{i}$.
\end{enumerate}
Over many generations, the population migrates toward low objective values. DE is stochastic, can escape shallow local minima better than single-start local methods, and is a natural second stage after nested sampling has already narrowed the search region. After \texttt{Nautilus}, we start DE from the weighted median of the final 100 posterior samples and search within bounds set by the 2.1\%--97.9\% weighted quantiles of the same samples rather than the full prior box. This starting point is distinct from the full-posterior marginal medians shown in the corner plots (Figures~\ref{fig:hubble_corner} and~\ref{fig:pearl_corner}). In this work, we further polish the DE result by performing nonlinear optimization using the best DE point as a starting guess for the L-BFGS-B optimizer available through the \bashinline{scipy.optimize} module's \bashinline{minimize} method. 

\section{Applying to Telescope Models}
\subsection{Hubble secondary alignment}\label{sec:hubble}

\begin{figure}
    \centering
    \includegraphics[width=\columnwidth]{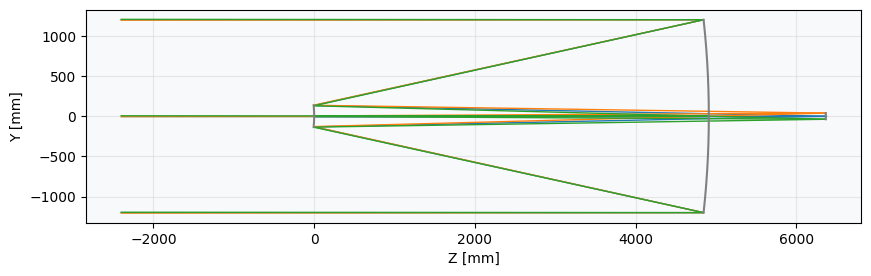}
    \caption{Optiland ray trace of the Hubble model. Likelihood evaluation uses circular-Bessel coefficients at five field coordinates (see text).}
    \label{fig:hubble}
\end{figure}

As an example of a 2-mirror system, we model a Ritchey--Chr{\'e}tien-style Hubble Space Telescope in Optiland (Figure~\ref{fig:hubble}; EPD\,$=$\,2400\,mm, $\lambda=0.55\,\mu\mathrm{m}$) with misalignments applied only to the secondary mirror. The five degrees of freedom are: decenter $sdx$, $sdy$ and despace $sdz$ (microns), and tilts $stx$, $sty$ (degree). Normalized field coordinates are relative to a maximum field angle of $0.038^{\circ}$.

PSFs are computed with \bashinline{MMDFTPSF} at \bashinline{grid_size=64}, cropped to $64\times64$ pixels, and projected onto 15 circular-Bessel basis functions. Observations are noise free. Synthetic truth coefficients are stored at five normalized fields: $(0,0)$, $(1,0)$, $(0,1)$, $(1,1)$, and $(-1,-1)$. For each field, a $15\times15$ covariance matrix is estimated from $10^4$ Monte Carlo draws of $\boldsymbol{\theta}$ under zero-mean Gaussian perturbations ($\sigma=100\,\mu\mathrm{m}$ on translations, $\sigma=0.167$ degrees on tilts), with each draw rounded to two decimal places.

Uniform priors on the unit cube map to $\pm200\,\mu\mathrm{m}$ on $(sdx,sdy,sdz)$ and $\pm0.167$ degrees on $(stx,sty)$; samples from the prior transform are rounded to $0.01\,\mu\mathrm{m}$ and $0.01$ degrees. Nautilus runs with \bashinline{n_dim=5}. In our completed benchmark, the sampler performed 42\,800 likelihood evaluations with $\ln\mathcal{Z}\approx +107.5$ and $N_{\mathrm{eff}}\approx 1.43\times 10^4$. The weighted median of the last 100 posterior samples provides the starting point for DE.

The DE objective minimizes $\sum_{k=1}^{15}[c_k(\boldsymbol{\theta})-c_{k,\mathrm{obs}}]^2$ at the on-axis field only. Multi-field terms are not included in the DE stage. Nautilus constrains $sdz$ and the tilts but leaves $sdx$ and $sdy$ broad (Section~\ref{sec:hubble_results}); in the noise-free case, the on-axis coefficients alone are sufficient for DE to recover all five parameters within the posterior-derived bounds.

\begin{figure}
    \centering
    \includegraphics[width=\columnwidth]{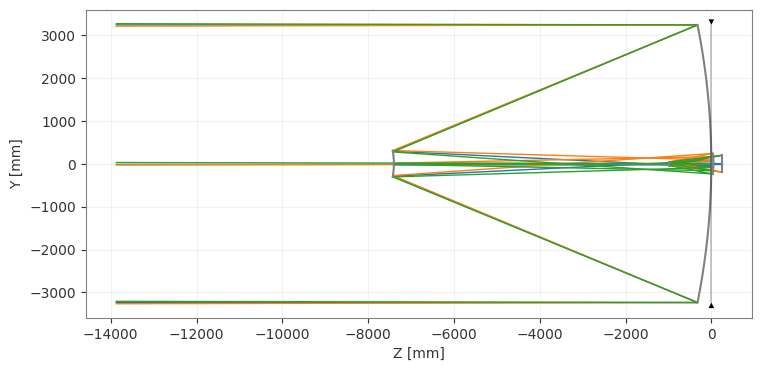}
    \caption{Raytrace of the TMA model in Optiland. Multi-field PSF features used in inference are evaluated at ten field coordinates across the focal plane (see text).}
    \label{fig:pearltma}
\end{figure}

\begin{figure}
    \centering
    \includegraphics[width=\columnwidth]{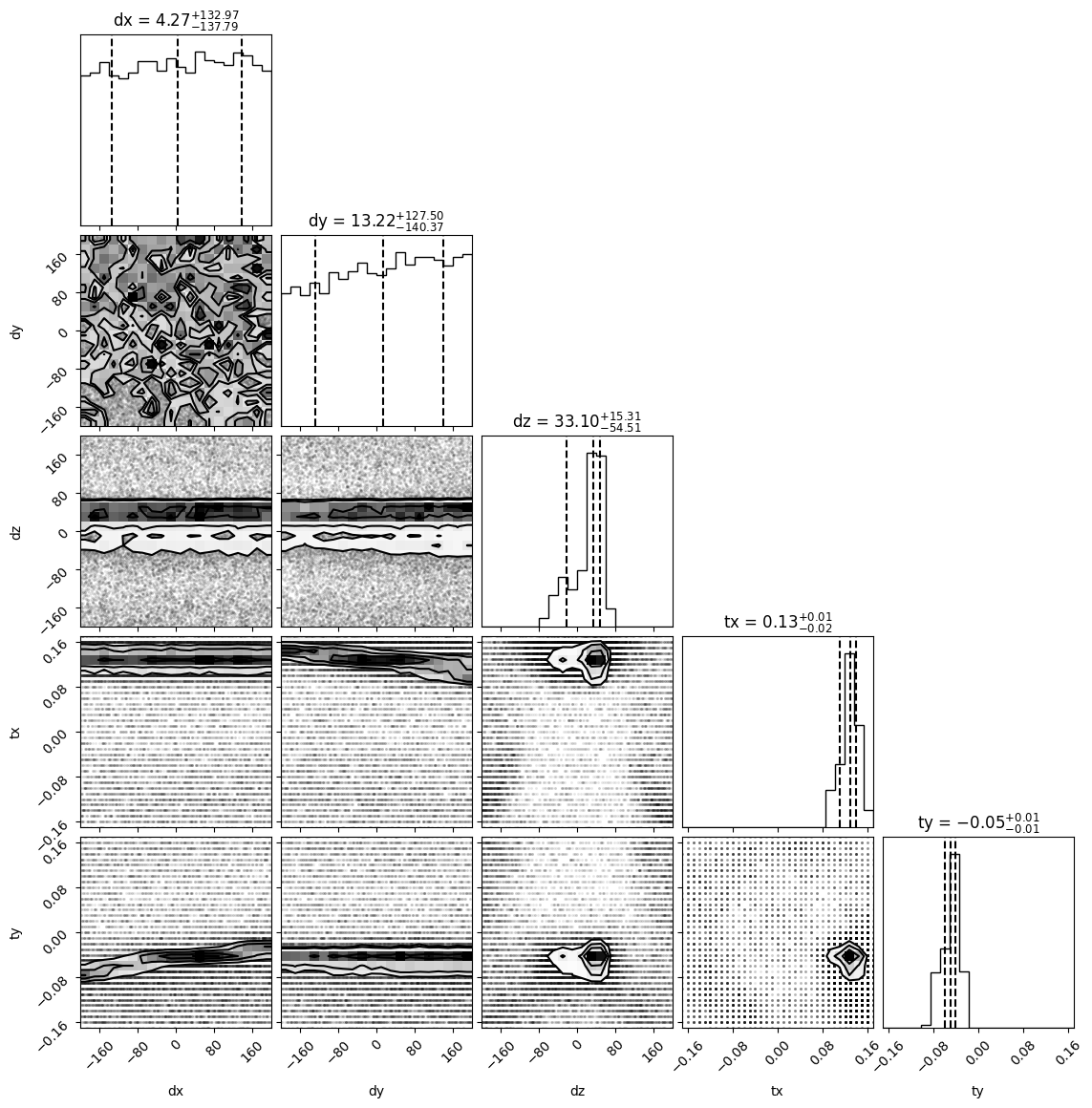}
    \caption{Nautilus posterior corner plot for the five secondary-misalignment parameters of the Hubble model (axes labelled $dx$, $dy$, $dz$, $tx$, $ty$ correspond to $sdx$, $sdy$, $sdz$, $stx$, $sty$). Titles give the full-posterior marginal medians.}
    \label{fig:hubble_corner}
\end{figure}

\begin{figure*}
    \centering
    \includegraphics[width=\textwidth]{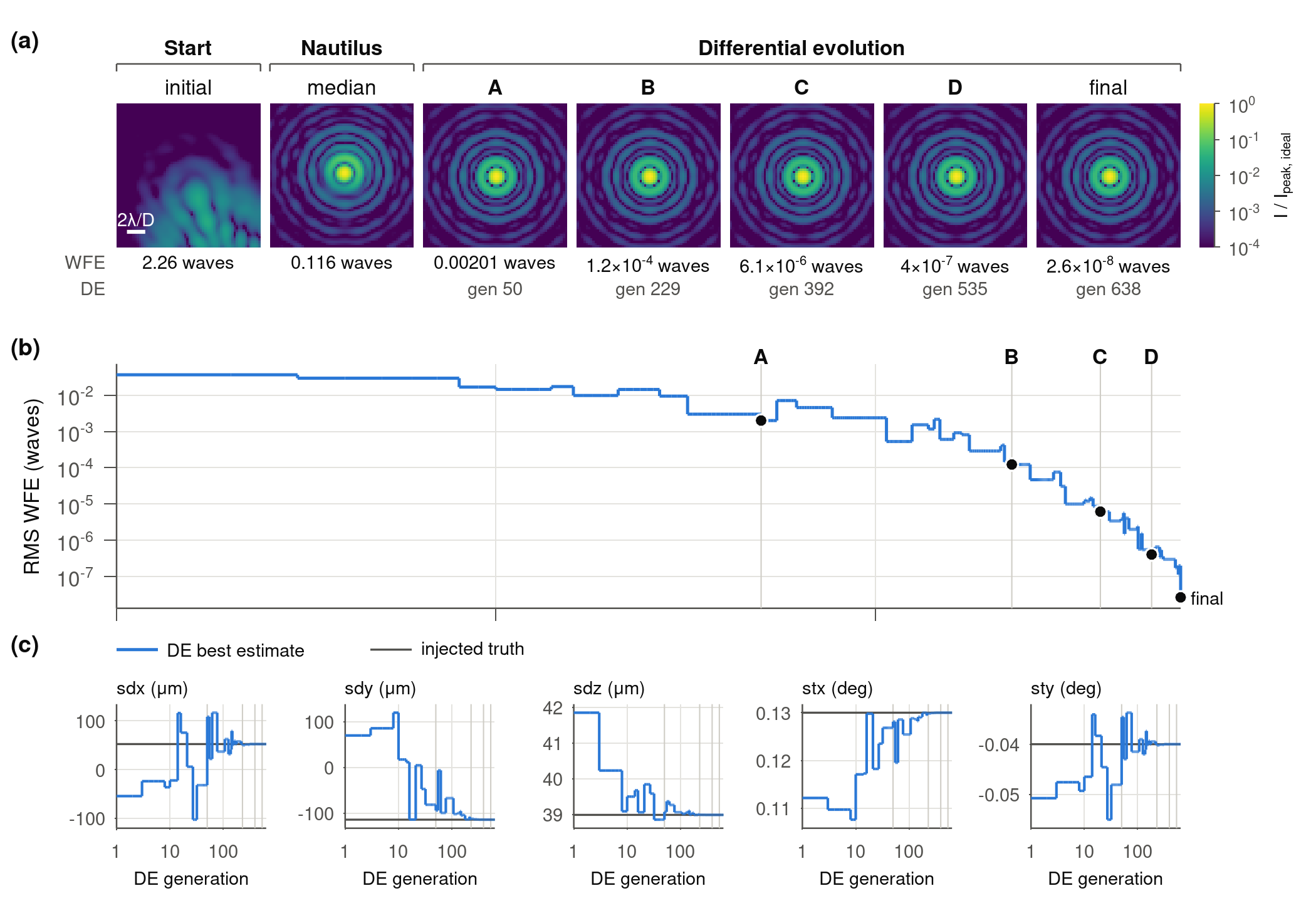}
    \caption{Hubble benchmark: on-axis PSF and RMS wavefront error (WFE; the on-axis OPD RMS quoted in the text) as the alignment is refined. (a) PSFs for the uncorrected system, the Nautilus starting point (median of the final 100 posterior samples), intermediate DE solutions and the final DE solution. (b) RMS WFE of the DE best estimate versus generation. (c) DE best estimate of each secondary-mirror parameter versus generation, with the injected truth.}
    \label{fig:psf_prog}
\end{figure*}

\begin{figure*}
    \centering
    \includegraphics[width=\textwidth]{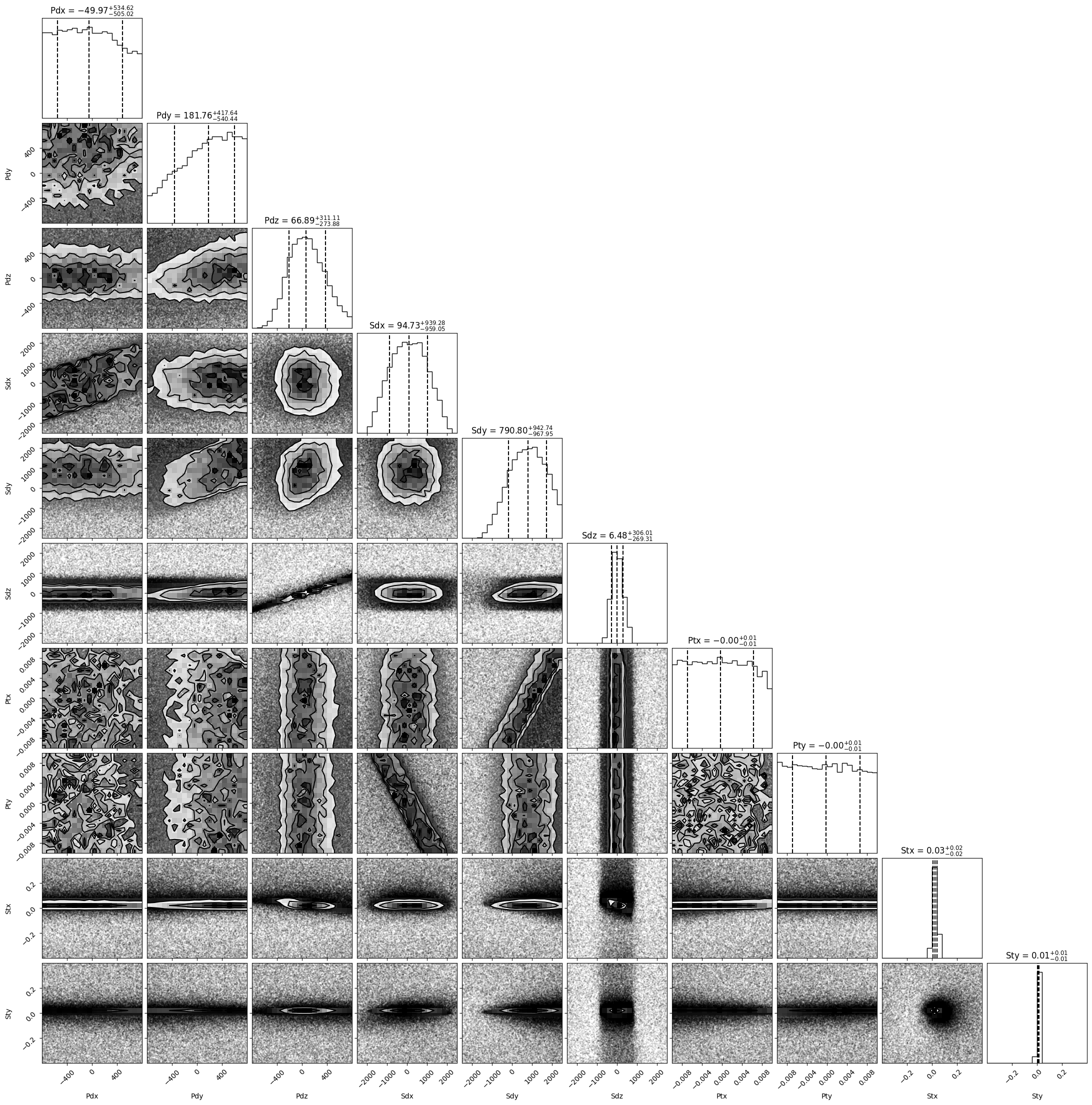}
    \caption{Nautilus posterior corner plot for the ten primary/secondary misalignment parameters of the TMA model.}
    \label{fig:pearl_corner}
\end{figure*}

\subsection{Three mirror anastigmat}\label{sec:pearl}

We apply the same inference pipeline to a compact three-mirror anastigmat (TMA) based on the optical design described in \citet{PearlTMA}, implemented in Optiland as a four-mirror prescription (Figure~\ref{fig:pearltma}) with entrance-pupil diameter EPD\,$=$\,6471.26\,mm and monochromatic wavelength $\lambda = 0.65\,\mu\mathrm{m}$. Unlike the Hubble case, misalignments are applied to both the primary and secondary mirrors: ten rigid-body degrees of freedom parameterize decenter ($x,y$), despace ($z$), and tilt ($x,y$) on each mirror. Translations are specified in microns and tilts are in degrees. Normalized field coordinates are relative to a maximum field angle of $0.285^{\circ}$.

For each trial alignment vector $\boldsymbol{\theta}$, we build the TMA model and propagate the field with \bashinline{MMDFTPSF} using 512 rays, and retain a central $256\times256$ pixel intensity patch. The patch is projected onto the first 11 Zernike modes, yielding an 11-element coefficient vector $\mathbf{c}_f(\boldsymbol{\theta})$ at each field point $f$. The synthetic observations $\mathbf{c}_{f,\mathrm{obs}}$ are computed from PSFs with additive Gaussian noise ($\sigma=0.01$ per pixel, in PSF intensity units), while forward-model PSFs in the likelihood are noise free. Each covariance $\boldsymbol{\Sigma}_f$ is estimated from $10^4$ noise-free Monte Carlo draws with $\sigma=250\,\mu\mathrm{m}$ on $(pdx,pdy)$, $50\,\mu\mathrm{m}$ on $pdz$, $800\,\mu\mathrm{m}$ on $(sdx,sdy)$, $100\,\mu\mathrm{m}$ on $sdz$, $0.003$ degrees on $(ptx,pty)$, and $0.13$ degrees on $(stx,sty)$, with each draw rounded to two decimal places.

The prior transform maps uniform unit-cube samples to independent uniform boxes: primary decenters $(pdx,pdy)$ and despace $pdz \in [-800,800]\,\mu\mathrm{m}$, secondary decenters $(sdx,sdy)$ and despace $sdz \in [-2500,2500]\,\mu\mathrm{m}$, primary tilts $\in [-0.01,0.01]$ degrees, and secondary tilts $\in [-0.4,0.4]$ degrees. The weighted median $\mathbf{x}_0$ and the 2.1\%--97.9\% weighted quantiles of the final 100 posterior samples initialize and bound differential evolution. The DE objective is the unweighted sum of squared coefficient residuals over all ten fields, $\sum_{f\in\mathcal{F}}\sum_{k=1}^{11}[c_{f,k}(\boldsymbol{\theta})-c_{f,k,\mathrm{obs}}]^2$, minimized with \bashinline{tol=1e-10}; the run reached the limit of 2000 generations (final objective $2.2\times10^{-7}$) before L-BFGS-B polishing.

Recovered alignments $\mathbf{c}$ are validated by constructing a residual telescope with misalignments $(\boldsymbol{\theta}_{\mathrm{true}} - \mathbf{c})$ and comparing on-axis PSF images and OPD maps against the nominal system via \bashinline{OPD} at field $(0,0)$. We report both per-component residuals $\boldsymbol{\theta}_{\mathrm{true}} - \mathbf{c}$ and the RMS of the OPD difference in waves.

Both studies use Optiland for ray tracing, Nautilus for nested sampling, and SciPy~\citep{Virtanen2020} differential evolution for final refinement; the TMA study uses POPPY~\citep{Perrin2012} for the Zernike basis.

\section{Results}

We evaluated both pipelines in simulation using analysis notebooks that record truth misalignments $\boldsymbol{\theta}_{\mathrm{true}}$, nested-sampling posteriors, DE solutions $\mathbf{c}$, and validation metrics. The Hubble benchmark projects PSFs onto circular-Bessel basis functions, while the TMA study uses Zernike polynomials from POPPY. The two bases weight the PSF very differently. Zernike polynomials are orthogonal on the unit disk and grow toward its edge, so projecting a PSF intensity onto the first few modes gives, in effect, low-order image moments: piston measures the total flux, tip and tilt the centroid, defocus the radial width, astigmatism the ellipticity, and coma the third-order asymmetry. Because these weights rise with radius and are nearly flat across a compact core, Zernike features respond well to broad, strongly aberrated PSFs but give heavy weight to the faint, noise-dominated wings and are largely blind to the ring structure of a nearly diffraction-limited PSF. The circular-Bessel functions, $J_{n+1}(2\pi\rho)\cos(m\phi)/\rho$ and $J_{n+1}(2\pi\rho)\sin(m\phi)/\rho$, are the Fourier transforms of the Zernike polynomials~\citep{Noll1976}, so their radial structure follows the diffraction rings of a circular pupil (the lowest-order term is the Airy amplitude) and their weight decays away from the core. Circular-Bessel features should therefore be more sensitive to small misalignments and less affected by noise in the wings, at the cost of a radial scale that must be matched to $\lambda/D$ in detector pixels and a smaller capture range for large misalignments that scatter light far from the core. A Zernike basis may therefore suit a coarse search from a strongly misaligned state, and a circular-Bessel basis the final refinement; a controlled comparison of the two on the same model and noise level is left to future work.

\subsection{Hubble benchmark}\label{sec:hubble_results}

One noise-free realization was injected with translations $(sdx,sdy,sdz)=(52.22,\,-113.86,\,38.99)\,\mu\mathrm{m}$ and tilts $(stx,sty)=(0.13,\,-0.04)$ degrees. Nautilus ($\ln\mathcal{Z}\approx +107.5$) tightly constrained both tilts and, more loosely, $sdz$, but left $sdx$ and $sdy$ unconstrained: their marginals in Figure~\ref{fig:hubble_corner} are close to flat across the $\pm200\,\mu\mathrm{m}$ prior. The weighted median of the final 100 posterior samples, used to start DE, was $(131.13,\,56.64,\,40.87,\,0.12,\,-0.03)$, and the corresponding 2.1\%--97.9\% DE bounds on $sdx$ and $sdy$ each span $\approx 310$--$320\,\mu\mathrm{m}$. Differential evolution converged to $\mathbf{c}=(52.22,\,-113.86,\,38.99,\,0.130,\,-0.040)$ with final objective $\lesssim 10^{-15}$, yielding sub-$0.01\,\mu\mathrm{m}$ residuals on all translations (Table~\ref{tab:hubble}).

\begin{table}
\centering
\caption{Hubble secondary: injected truth and DE-recovered misalignment (noise-free benchmark). Tilts are in degrees.}
\label{tab:hubble}
\begin{tabular}{llll}
\hline
\textbf{Parameter} & \textbf{True value} & \textbf{Recovered (DE)} & \textbf{Units} \\
\hline
sdx & $52.22$  & $52.22$  & $\mu$m \\
sdy & $-113.86$ & $-113.86$ & $\mu$m \\
sdz & $38.99$  & $38.99$  & $\mu$m \\
stx & $0.13$   & $0.13$   & degree \\
sty & $-0.04$  & $-0.04$  & degree \\
\hline
\end{tabular}
\end{table}

With no correction, on-axis OPD RMS is $\sim 2.26$ waves. Using the DE starting point (the median of the final 100 posterior samples) alone reduces this to $\sim 0.12$ waves. After DE refinement, on-axis OPD RMS falls to $\sim 3\times 10^{-8}$ waves. Figure~\ref{fig:hubble_corner} shows posterior marginals for all five secondary misalignments, and Figure~\ref{fig:psf_prog} shows the on-axis PSF and RMS wavefront error at each stage of the refinement.

\subsection{TMA benchmark}

The TMA study uses the noisy forward model described in Sec.~\ref{sec:pearl}. Ten-dimensional nested sampling is computationally expensive; posterior medians need not coincide with injected truth when the likelihood contains flat directions. The posterior marginals in Figure~\ref{fig:pearl_corner} suggest partial degeneracies in the ten-dimensional problem. The primary decenter $pdx$ and both primary tilts are close to flat across their prior ranges, the secondary decenters remain broad, and several pairs are strongly correlated, notably $pdz$--$sdz$, $ptx$--$sdy$ and $pty$--$sdx$. The tilt--decenter correlations have a familiar optical origin: a tilt of one mirror about the $x$ ($y$) axis and a decenter of the other along $y$ ($x$) both introduce coma that is nearly constant across the field, so one can largely cancel the coma introduced by the other. The same mechanism made secondary-mirror misalignment on JWST detectable only through field-dependent aberrations (Section~\ref{sec:intro}): a compensating motion of another mirror cancels the field-constant error and leaves a field-dependent residual, which is why our likelihood combines several field points. The $pdz$--$sdz$ correlation similarly reflects the trade-off between the two despaces in setting focus. Only the secondary tilts are tightly constrained. Distinct alignment states may therefore produce similar Zernike vectors, and the likelihood alone may not determine every parameter uniquely.

\begin{figure*}
    \centering
    \includegraphics[width=\textwidth]{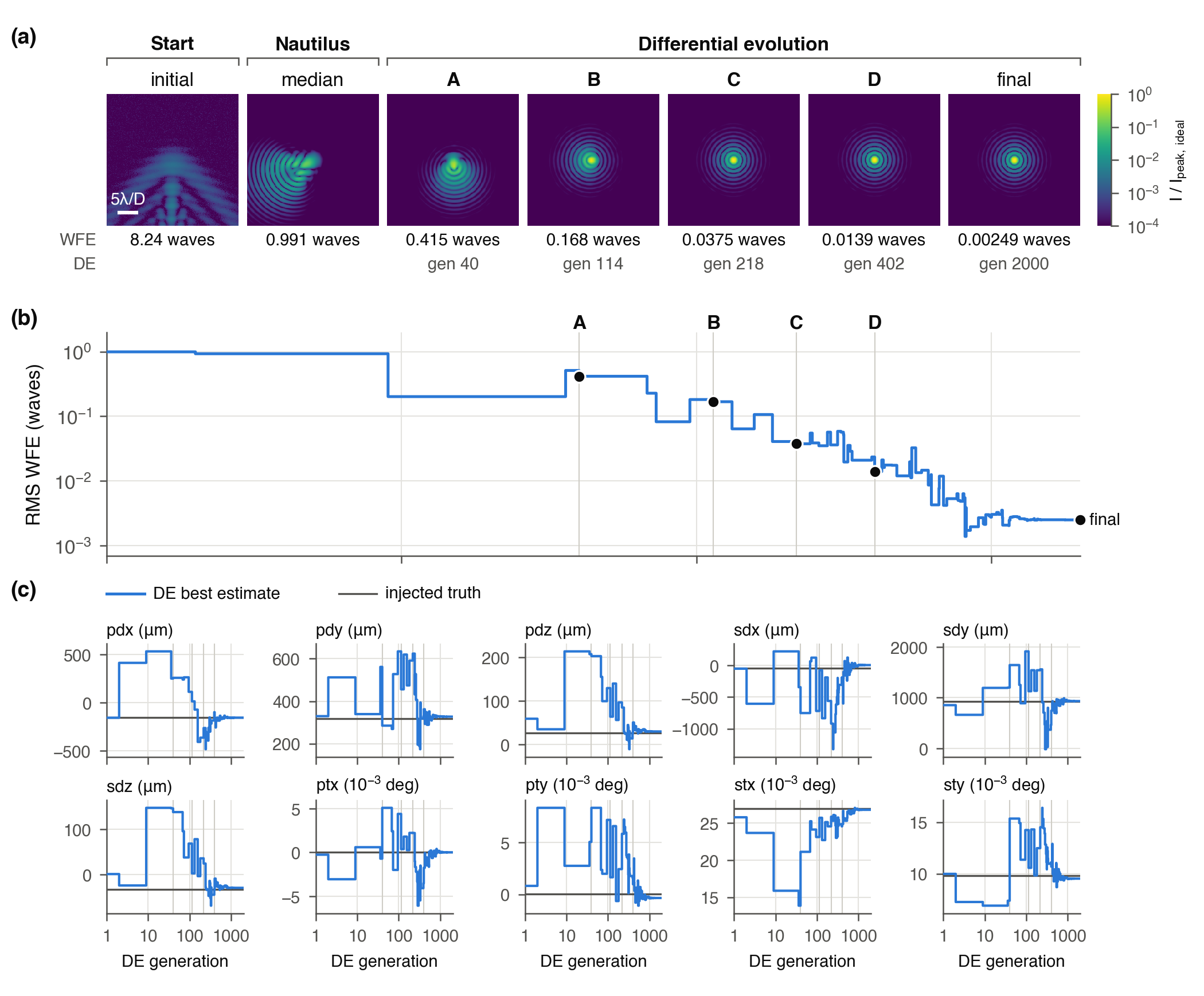}
    \caption{TMA benchmark: (a) On-axis PSFs at $\lambda = 0.65\,\mu\mathrm{m}$: the observed misaligned PSF (Gaussian noise, $\sigma = 0.01\%$ of the ideal peak), the Nautilus starting point (median of the final 100 posterior samples), during differential evolution (DE) at generations 40, 114, 218 and 402 (A--D), and after 2000 generations (final). Intensity is relative to the aligned system's peak, so each panel's peak equals its Strehl ratio (log stretch, floor $10^{-4}$). Values are RMS wavefront error in waves, the RMS of the OPD difference from the nominal design. (b) RMS WFE of the best DE member at each generation; letters mark the PSFs in (a). (c) Best-member value of each alignment parameter at each generation (blue) and the injected truth (gray).}
    \label{fig:tma_psf_prog}
\end{figure*}

Nautilus performed 93\,200 likelihood evaluations with $\ln\mathcal{Z}\approx +484.7$ and $N_{\mathrm{eff}}\approx 2.04\times 10^4$. With no correction, on-axis OPD RMS is $\sim 8.24$ waves, and the DE starting point $\mathbf{x}_0$ (the median of the final 100 posterior samples) alone reduces this to $\sim 0.99$ waves. Table~\ref{tab:pearl} reports the truth and DE-refined alignment states: the refinement reduces on-axis OPD RMS to $\sim 2\times 10^{-3}$ waves while leaving residuals of up to $\sim 16.5\,\mu\mathrm{m}$ on the translations (largest on secondary decenter $sdy$). Primary tilts are near zero at truth and remain poorly determined relative to translations. Figure~\ref{fig:tma_psf_prog} shows the on-axis PSF and RMS wavefront error at each stage of the refinement.

\begin{table}
\centering
\caption{TMA: injected truth and DE-recovered alignment.}
\label{tab:pearl}
\begin{tabular}{llll}
\hline
\textbf{Parameter} & \textbf{True value} & \textbf{Recovered} & \textbf{Units} \\
\hline
pdx  & $-152.75$ & $-159.43$ & $\mu$m \\
pdy  & $316.91$  & $309.97$  & $\mu$m \\
pdz  & $26.63$   & $24.77$   & $\mu$m \\
sdx  & $-42.31$  & $-38.01$  & $\mu$m \\
sdy  & $923.33$  & $939.85$  & $\mu$m \\
sdz  & $-32.75$  & $-34.56$ & $\mu$m \\
ptx  & $2.98\times 10^{-5}$ & $2.22\times 10^{-4}$ & degree \\
pty  & $1.61\times 10^{-5}$ & $6.80\times 10^{-5}$  & degree \\
stx  & $0.0269$  & $0.0272$  & degree \\
sty  & $0.0098$  & $0.0098$  & degree \\
\hline
\end{tabular}
\end{table}

\section{Conclusion}

We presented a Bayesian pipeline for inferring telescope rigid body misalignment from simulated multi-field observations for optics with known prescriptions. Nested sampling with Nautilus provides global posterior exploration and credible regions; differential evolution delivers a high-precision point estimate seeded by the median of the final posterior samples. In a noise-free Hubble secondary-mirror benchmark, the method recovers five degrees of freedom to $\lesssim 0.01\,\mu\mathrm{m}$ and reduces on-axis OPD error from $\sim 2.3$ waves to $\sim 3\times 10^{-8}$ waves. In a ten-degree-of-freedom TMA study with noisy PSFs, DE recovered the translations to $\lesssim 17\,\mu\mathrm{m}$ and reduced on-axis OPD error from $\sim 8.24$ waves to $\sim 2\times 10^{-3}$ waves. This approach, while computationally intensive allows measurement of the state of an optical system with a single measurement of a wide-field imaging instrument as is typical of modern TMA observatories  and is expected to enable deterministic alignment of optical systems.

Future work includes extending the pipeline to chromatic and realistic noise models and applying the pipeline to laboratory or on-sky PSF campaigns relevant to Habitable Worlds Observatory class TMA designs.

\section*{Acknowledgements}

J.N.A was supported by NASA through the NASA Hubble Fellowship grant \#HST-HF2-51547.001-A awarded by the Space Telescope Science Institute, which is operated by the Association of Universities for Research in Astronomy. The authors also thank Tim Eifler and Noah Franz for insights related to Nested sampling and Kevin Derby and Solvay Blomquist for general advice. 

Portions of this research were supported by generous anonymous philanthropic donations to the Steward Observatory of the College of Science at the University of Arizona.
\section*{Data Availability}

The Jupyter notebooks containing code, analysis, and figures used in this work are available on Zenodo: \url{https://doi.org/10.5281/zenodo.23627435}. 



\bibliographystyle{rasti}
\bibliography{example} 



\bsp	
\label{lastpage}
\end{document}